\documentclass[twocolumn,preprintnumbers,amsmath,amssymb]{revtex4}
\usepackage{graphicx}
\usepackage{dcolumn}
\usepackage{amsmath}
\usepackage{bm}

\begin{document}

\title{Membrane amplitude and triaxial stress in twisted bilayer graphene
deciphered using first-principles directed elasticity theory and scanning tunneling microscopy}

\author{M. Neek-Amal$^1$, P. Xu$^2$, D. Qi$^2$, P. M. Thibado$^2$, L. O. Nyakiti$^3$, V. D. Wheeler$^4$, R. L. Myers-Ward$^4$, C. R. Eddy,
Jr.$^4$, D. K. Gaskill$^4$, and F.M. Peeters$^1$}
\affiliation{$^1$Department of Physics, University of Antwerpen, Groenenborgerlaan 171, B-2020 Antwerpen, Belgium\\
$^2$Department of Physics, University of Arkansas, USA\\
$^3$U.S. Naval Research Laboratory, Washington, District of Columbia
20375, USA\\\\
$^4$Departments of Marine Engineering, Material Science and
Engineering, Texas A$\&$M University, College Station TX, 77843 USA
}
\date{\today}

\begin{abstract}

Twisted graphene layers produce a moir\'e pattern (MP) structure
with a predetermined wavelength for  given twist angle. However,
predicting the membrane corrugation amplitude for any angle other
than pure AB-stacked or AA-stacked graphene is impossible using
first-principles density functional theory (DFT) due to the large
supercell. Here, within elasticity theory we define the MP structure
as the minimum energy configuration, thereby leaving the height
amplitude as the only unknown parameter. The latter is determined
from DFT calculations for AB and AA stacked bilayer graphene in
order to eliminate all fitting parameters. Excellent agreement with
scanning tunneling microscopy (STM) results across multiple
substrates is reported as function of twist angle.

\end{abstract}
\maketitle
\section{introduction}
 The electronic properties of twisted stacked
graphene layers have been the focus of numerous
studies~\cite{Lopes}. The periodic potential of the interacting
substrate is the source of a new set of Dirac points in the energy
spectrum of graphene~\cite{6}. Also, the van Hove singularity is
found to shift with twist angle~\cite{prl,eva}. For large angles the
graphene layers behave like isolated sheets, while for small angles
the new Dirac cones result in two van Hove
singularities.~\cite{zhao}.

Early experimental studies of multi-layer twisted graphene using
scanning tunneling microscopy (STM) found a moir\'e pattern (MP)
structure ~\cite{eva}. Such a MP results in an additional
corrugation as compared to the untwisted case. The most prominent
examples have come from epitaxial graphene grown on
SiC~\cite{PRL2008,science2009,PHYSREVB2009}. From those experimental
data, a simple analytic expression for the wavelength of the
superstructure was quickly discovered and provided a clear picture
of the mechanism as well as the responsible twist angle. Much more
difficult, however, is predicting the corrugation amplitude, and so
far a simple analytic expression for this does not exist.
 Theoretical studies about the
height deformation of the twisted bilayer graphene are difficult.
This is because the large size (e.g. up to 10 nm in size) MP unit
cell makes ab-initio calculations infeasible. Only in certain
limiting cases the size of the unit cell is sufficiently small that
ab-initio calculations are possible~\cite{SACH}. When DFT results
can be obtained they set the standard for all other approaches.
Consequently, it is best to parameterize any new approach such that
it agrees with the DFT results in certain limits
~\cite{vdw,scientificreports,spanu}. Nevertheless, there exist
alternative methods that show promise using classical interatomic
potentials~\cite{PCCP}.

Here we present an analytical approach for the height deformation of
twisted bilayer graphene without using any fitting parameters and
assuming only that the experimentally observed MP structure is the
minimum energy configuration. We show that the deformation of the
top graphene layer, due to the van der Waals interaction, is
affected by the MP pattern. These deformations result in strain
which subsequently leads to three-fold symmetry in the curvature and
an induced pseudo-magnetic field. We also report excellent agreement
with scanning tunneling microscope measurements that we acquired
from various multi-layer graphene samples.

\section{The samples and STM experiments}
  Multiple epitaxial
graphene samples grown on various miscut (i.e., non-polar m-plane
and a-plane surfaces) n+ 6H-SiC substrates (measuring 16 mm $\times$
16 mm, Aymont Technology) were used for this study. Growth was
carried out in a commercially available hot-wall Aixtron VP508
chemical vapor deposition reactor. Prior to graphene growth, both
SiC substrates were etched in situ in a 100 mbar
 H$_2$ ambient at either 1520$^o$C or
1560$^o$C for 50 min. After etching, the ambient condition was
switched to Ar, followed by a temperature ramp to 1620$^o$C. The
graphene synthesis process was then conducted for 15 minutes up to
60 minutes under a flowing Ar environment of 20 standard liters per
min. at 100 mbar, with a substrate temperature still at 1620$^o$C.
The post-growth morphology was characterized using atomic force
microscopy and the multi-layer graphene coverage was confirmed using
Raman spectroscopy. After these characterizations, constant-current
filled-state STM images were obtained using an Omicron
ultrahigh-vacuum (base pressure is 10$^-$10 mbar), low temperature
model STM operated at room temperature. The samples were mounted
with silver paint onto a flat tantalum sample plate and transferred
through a load-lock into the STM chamber where it was electrically
grounded. STM tips were electrochemically etched from 0.25 mm
diameter polycrystalline tungsten wire via a custom double lamella
method with an automatic gravity-switch cutoff. After etching, the
tips were gently rinsed with distilled water; briefly dipped in a
concentrated hydrofluoric acid solution to remove surface oxides,
and then transferred into the STM chamber. Additional experimental
details are provided elsewhere~\cite{more}.

\section{The model}
\subsection{Minimum energy configuration}
 For a given twist angle $\theta$ between
two graphene layers, the top sheet is attracted to the bottom sheet
due to van der Waals (vdW) interaction. The zero lattice  mismatch
between the honeycomb lattice structures of the two graphene layers
leads to an infinite
 moir\'e wavelength $L$ when the two layers have
either AB- or AA-stacking. This is because
$L=\frac{\sqrt{3}a_{0}}{2\,sin(\theta/2)}$ where $a_{0}=1.42\,\AA$
is the carbon-carbon bond length, $\theta$ is the disorientation
angle with respect to AB-stacking having $\theta=0$, and AA-stacking
corresponds to $\theta=\pi/3$. In general the commensurate rotation
where a B' atom from the  top layer is  directly above the A atom in
the bottom layer,  is moved by the rotation to a position formerly
occupied by an atom of the same kind, can be obtained from
\begin{equation}
\theta_n=\cos^{-1}[\frac{3n^2+3n+1/2}{3n^2+3n+1}],~~n=0,1,2,..
\end{equation}

\begin{figure*}
\begin{center}
\includegraphics[width=1\linewidth]{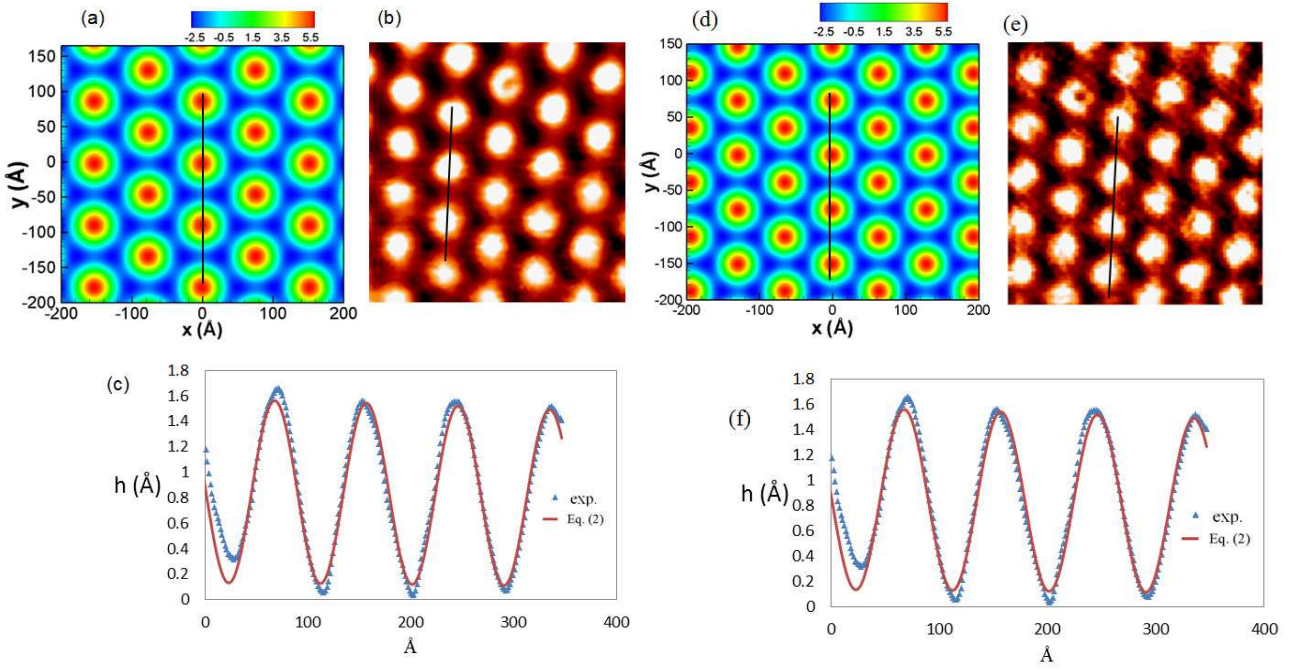}
 \caption{ Height deformation  of a graphene sheet over graphene: (a,d) are the results from Eq.~(\ref{h2})
  and corresponding filled-state (0.05 V), constant-current (1.0 nA) STM images (b,e) for twist angles $\theta=1.59^{o}$(a,b,c) and
 $\theta=1.88^{o}$(d,e,f). In (c,f) we show two  cross sections along indicated solid
  black lines in the top figures. }\label{figh}
\end{center}
\end{figure*}

For twist angle $\theta$ defined with respect to the $x$-axis (taken
along the zigzag chain direction of graphene), we define the
out-of-plane deformation of the lattice as $h(\vec{r},\theta)$ where
$\vec{r}=(x,y)$. From experiment we know that the minimum energy
configuration for $h(\vec{r},\theta)$ is the MP structure and
depending on the preparation method different twist angles are
possible. Furthermore, from continuum elasticity theory
 the deformation of the membrane over a flat substrate is given by
 the solution of the following differential equation ~\cite{prbneek2012}:
\begin{equation}(k\nabla^4-\tau\nabla^2+v(\vec{r},\theta))Z(\vec{r},\theta)=0,\label{eq1}
\end{equation}
where $Z(\vec{r},\theta)$ is the height of the membrane at
$\vec{r}$, $k$ and $\tau$ correspond to the bending and stretching
modulus of graphene and $v(\vec{r},\theta)$ depends on the vdW
parameters between the two layers and is proportional to the Hamker
constant. The Fourier transform (FT) of the solution of
Eq.~(\ref{eq1}) must have six moir\'e pattern
vectors~\cite{4,6,APL2014,Wallbank}, i.e., $\vec{G}_{m}=
\Re_{\phi_m}\vec{G}_0$ with $m=0,1,..5$ where
$\vec{G}_0=(1-\Re_{\theta})(0,2\kappa)$ with
$\kappa=\frac{2\pi}{3a_{0}}$ and $\Re_{\phi_m}$ (and $\Re_{\theta}$)
is the rotation matrix about the $z$-axis over an angle
$\phi_m=\frac{2 \pi m}{6}$ (and $\theta$).

Therefore, for $\theta$$>$0 the height deformation of graphene  can
be written as
\begin{equation} h(\vec{r},\theta)=h_0 (\theta)
f(\vec{r},\theta)\end{equation}
 where the modulation function is
$f(\vec{r},\theta)=\sum_m e^{i \vec{G}_m.\vec{r}}$~\cite{Wallbank},
and $h_0(\theta)$ should be determined using microscopic information
(the zero reference height is taken to be the AB-stacking interlayer
position, i.e. $Z(\vec{r},\theta)=d_{AB}+h(\vec{r},\theta)$, and
corresponds to the minimum energy configuration). For a given twist
angle we can simplify the modulation function as

\begin{equation}
 h=2 h_0(\theta)\{\cos [\vec{r}. \vec{G_0} ]+2\cos
[\frac{\vec{r}. \vec{G_0}}{2}]\cos[\frac{\sqrt{3}}{2}|\vec{r}\times
\vec{G_0}|]\}.\label{h2}
\end{equation}

In order to better visualize and to compare it with experimental
data, we plot this function in Fig.~\ref{figh} for two typical twist
angles of 1.59$^o$ in (a) and 1.88$^o$ in (d).  A plethora of STM
images showing various MP (not all shown) from various substrates
were collected together in order to experimentally test the theory.
The two items we accurately measure from the STM images are the
average wavelength as well as the average amplitude of the MPs. From
the wavelength measurement we convert to twist angle using the
formula mentioned earlier. For the amplitude measurements, two
situations arise. When the amplitude of the MP is large, similar to
the STM images shown in Figs.~\ref{figh}(b,e), it is easy to
determine the amplitude from the height cross section plots similar to
the ones shown in Figs.~\ref{figh}(c,f). Here, we show two height
cross sections from Figs.~\ref{figh}(a,d) along the solid black
lines and compare them with our experimental results (symbols) from
Figs. ~\ref{figh}(b,e).

\begin{figure*}
\begin{center}
\includegraphics[width=0.95\linewidth]{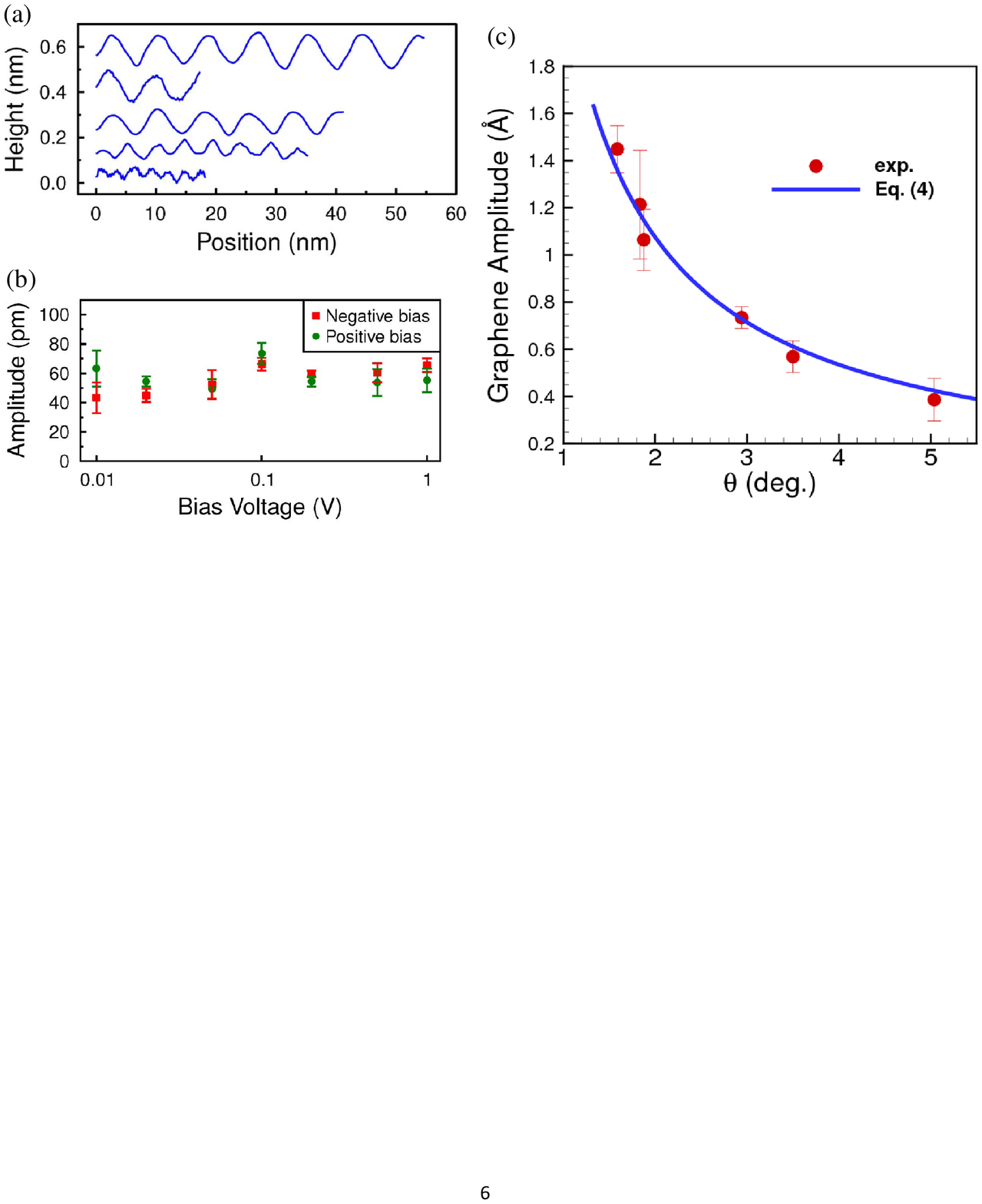}
 \caption{(a) Five typical constant-current height line profiles extracted from STM images acquired with a variety of tunneling setpoint conditions.
 (b) Constant current (0.5 nA) STM image derived data showing the average membrane amplitude versus bias voltage setpoint using a semi-log plot for twist angle 3.5$^o$.
 (c) STM image derived data (symbols) showing the average membrane amplitude versus twist angle.
 For the STM data we measure the wavelength ($\sim\pm$0.1 nm) and convert to twist angle. The solid line is the result
  of the presented theory, i.e. Eq.~(\ref{H02}).
\label{figexp}}
\end{center}
\end{figure*}

Notice that the height profile obtained from STM measurements give
us the total height which contains contributions from both
electronic and atomic corrugations ~\cite{PRBref}. The presented
theory in this work addresses only the atomic corrugations. The
electronic contributions depends on the used bias voltage and the
STM measurements conditions. In Ref.~\cite{prl}, a maximum of 50$\%$
of the total height was found to be due to the atomic corrugations.
However, when the amplitude is small the STM image shows a
superposition of the MP structure and the one due to the atomic
electronic corrugation. The electronic corrugation due to the
individual atoms is not part of the theory. For these STM images, we
measure the membrane amplitude by measuring the height change from
the top of the electronic corrugation at the top of the MP to the
top of the electronic corrugation at the bottom of the MP. For flat
graphene or graphite this height change gives zero. Note it is
possible that the electronic corrugation of the carbon atoms at the
top of the MP is slightly different when compared to the bottom of
the MP; however we expect this to be minute given the large
wavelengths and small amplitudes. The full collection of
experimental STM results for membrane height versus twist angle is
shown in Fig.~\ref{figexp} as symbols. Notice the excellent
agreement between theory and experiment, which supports the idea
that any electronic variation is small.

\section{Energetic consideration and strain tensor}

In order to have a real predictive theory we still need to calculate
$h_0(\theta)$. In order to do so, we first write the elastic energy as given
by
\begin{equation} E_{el}=\frac{1}{2} \int [k  (\nabla^2
h)^2+\lambda {\varepsilon_{\ell\ell}}^2 +2\mu
{\varepsilon_{ij}}^2]d\vec{r},\label{H0}
\end{equation}
where $k\simeq1.1$\,eV is the bending rigidity of graphene and
$\lambda$=3.5\,eV$\AA^{-2}$~and $\mu$=8\,eV$\AA^{-2}$~are the
Lam\'{e} coefficients. The elements of the strain tensor can be
found using $\epsilon_{\alpha\beta}= \frac{1}{2}(\partial_\beta
u_{\alpha}+\partial_\alpha u_{\beta})+\frac{1}{2}\partial_{\alpha} h
\partial_{\beta}h$:

\begin{widetext}
\begin{eqnarray*}
\\&&
\varepsilon_{xx}=4h_0^2
(\vec{G}_0.\vec{\chi})^2,\varepsilon_{yy}=4h_0^2
|\vec{G}_0\times\vec{\chi}|^2,\\&&\varepsilon_{xy}=4h_0^2
(\vec{G}_0.\vec{\chi})|\vec{G}_0\times\vec{\chi}|\\&&
\vec{\chi}=\{\sin [\frac{\vec{r}. \vec{G_0} }{2}](2\cos
[\frac{\vec{r}. \vec{G_0}
}{2}]+\cos[\frac{\sqrt{3}}{2}|\vec{r}\times
\vec{G_0}|]),\sqrt{3}\sin[\frac{\sqrt{3}}{2}|\vec{r}\times
\vec{G_0}|]\cos[\frac{\vec{r}. \vec{G_0}}{2}]\}.\label{strain}
\end{eqnarray*}
\end{widetext}

For the in-plane components (first term in parenthesis) we assume
that the top layer is in its minimum energy configuration (MP
structure) and the coordinate ($\vec{r}$) in our analysis is written
in the deformed system, thus we do not add them to the out-of-plane
components of the strain tensor~(see appendix). Diagonalising the
strain tensor gives the principal axis with eigenvalues
\begin{equation}\epsilon_{\pm}=\frac{1}{2}[\epsilon_{ll}\pm|\vec{A}|],
\end{equation}
 where $\vec{A}$ is the gauge vector corresponding to the lattice
deformation~\cite{geim}. Surprisingly, we found that  $\epsilon_-=0$
(since $\epsilon_{xx}\epsilon_{yy}=\epsilon_{xy}^2$) and that
$\epsilon_+=|\vec{A}|$ has MP symmetry. Since the eigenvalue
$\epsilon_-=0$, we conclude that the stress along the corresponding
eigenvector results in no lattice deformation. These principle
directions correspond to the most tensile and compression directions
in graphene, i.e.
\begin{equation}
\Phi_+=\frac{1}{2}tan^{-1}(\frac{-A_y}{A_x}),~~~~\Phi_-=\frac{1}{2}tan^{-1}(\frac{A_x}{A_y}).
\end{equation}

 Solution of the integrals in Eq.~(\ref{H0}) for a
given $\theta$ can be simplified to
\begin{equation} E_{el}=h_0^2(\theta)[
g_b(\theta)+h_0^2(\theta)g_s(\theta)],
\end{equation} where we found $g_i(\theta)$ by numerical integration ($a_i, b_i$ are also fitting parameters)
over the corresponding MP unit cell which has the following polynomial dependence:
\begin{equation}g_{b,s}(\theta)\cong(a_{b,s}+b_{b,s}\theta^2)^2.\end{equation}

In our recent work~\cite{APL2014}, we presented an atomistic
simulation showing that the local vdW energy stored between two
layers also exhibits a moir\'{e} pattern structure. Here using the
latter idea we write the binding energy as
\begin{equation}
E_{bin}=E_{AB}-\delta E
(1-\frac{\eta(\theta)}{\eta(0)}),\label{Hbin}
\end{equation}
where $E_{AB}\sim$\,50-60\,meV/atom is the binding energy between
two graphene layers in AB-stacking and $\delta
E=E_{AB}-E_{AA}(\approx\,13-15$ meV/atom as found from DFT in
Ref.~\cite{scientificreports}. \textbf{The binding energy varies
with interlayer distance as $d^{-4}$, but here we only model its
variation with $\theta$ and the in-plane coordinates because we are
only interested in the change in height, and not in its absolute
position.} Notice how this parameterization incorporates the known
DFT results. Note that the bright feature in all moir\'{e} patterns
is where we have local AA stacking of graphene. We can understand
this by realizing that in between two adjacent AA stacks there is a
low energy AB (i.e., Bernal) stacked region. Since carbon atoms in
an AA stack have higher energy as compared to the one for the AB
stack we expect larger amplitude in the AA stacked region, i.e. the
AB stacked planes are  closer together as compared to the AA stacked
planes.

We introduce the function $\eta(\theta)$ in Eq.~(\ref{Hbin}) based
on the MP as
\begin{equation}
\eta(\theta)=\int f(\vec{r},\theta) d\vec{r},
\end{equation}
which expresses the spacial average of the modulation function over
graphene.
\section{height profile and pseudo-magnetic field}
In mechanical equilibrium, the binding energy is competing with the
bending energy (elastic energy), and we must have
\begin{equation}E_{el}=E_{bin}.\end{equation} Solution of the latter equation results in
the following dependence for $h_0(\theta)$
\begin{equation}
h_0^2(\theta)=-\frac{g}{2}+[\frac{g^2}{4}+\frac{4}{3a_{0}^2}\frac{(E_{AB}-\delta
E (1-\frac{\eta(\theta)}{\eta(0)}))}{g_s}]^{1/2},\label{H02}
\end{equation}
where $g=g_b/g_s$. Because $g_b/g_s\ll1$ we can approximate $h_0$ as

\begin{equation}
h_0\cong[\frac{4}{3a_{0}^2}\frac{(E_{AB}-\delta E
(1-\frac{\eta(\theta)}{\eta(0)}))}{(a_s+b_s\theta^2)^4}]^{1/4}\sim
\theta^{-1},.
\end{equation}
for  $\theta>1^{o}$. Notice that the difference between the maximum
and the minimum of $h$ in Eq. (\ref{h2}) is given by $\Delta
h=8\,h_0$. $(E_{bin}-E_{AB})/(-\delta E)$  approaches 1 when
$\theta\rightarrow \pi/3$. Our prediction resulting from
Eq.~(\ref{H02}) for the overall height of the membrane (i.e. 8$h_0$)
is shown versus twist angle in Fig.~\ref{figexp} and compared with
our experimental results.

We collected  a variety of STM images of various multi-layer
graphene moir\'{e} patterns (not all are shown) from a-plane and
m-plane SiC substrates grown under similar conditions. Five typical
line profiles extracted from these STM images and having varying
wavelength and amplitude are shown in Fig. 2(a). The line profiles
are ordered from top to bottom based on decreasing amplitude. Notice
the lowest line profile has, superimposed on it, an even smaller
amplitude and higher frequency signal. This is the electronic
corrugation of the carbon atoms, and it is worth pointing out how
small the electronic amplitude is when compared to the membrane
corrugation. The membrane corrugation persists when imaging the
moir\'{e} pattern through a range of normal bias voltage settings
($\pm$0.05 to $\pm$1.00 V) and tunneling current setpoints (0.05 to
1.00 nA). For example, when a moir\'{e} pattern with a wavelength of
~4 nm is repeatedly imaged while incrementally altering the bias
voltage from $\pm$0.01 V to $\pm$1.0 V with a tunneling current
setpoint of 0.5 nA we see only a small amplitude variation as shown
in a semi-log plot of Fig.~2(b). Within the error bars, the membrane
amplitude is relatively unchanged.

 A plot showing the membrane amplitude as a function of twist angle
is shown in Fig.~2(c). Even though it is possible that the
electronic amplitude is slightly different at the crest of the
membrane compared to the trough of the membrane, we believe this is
within the error bars of our results. Also, unlike the image
contract inversion STM data acquired from single crystal metal
surfaces~\cite{PRBref}, for twisted graphene on graphene/SiC we do
not see any significant height changes in the moir\'{e} pattern as
we vary the STM tunneling condition.

\begin{figure}
\begin{center}
\includegraphics[width=1.1\linewidth]{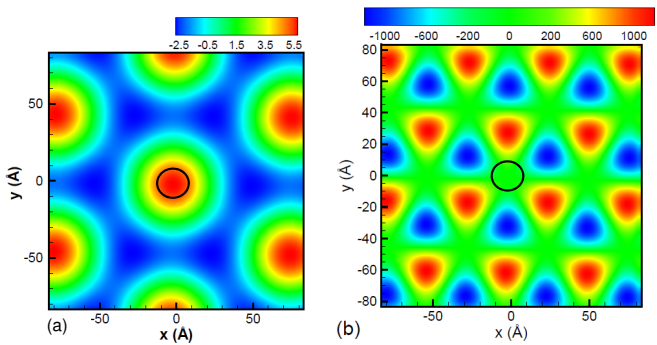}
 \caption{ Height deformation ($h/h_0$)  of graphene  over a graphene sheet, i.e. Eq.~(\ref{h2}),
  for twist angle  $\theta=1.59^{o}$(a) and corresponding
pseudo-magnetic field in units of $h_0^2$ (b), obtained from
  Eq.~(\ref{B}). The circle in (a), (b) indicate the regions with extreme height and
  zero magnetic field, respectively.
\label{figB}}
\end{center}
\end{figure}

It is also worthwhile to mention that the low energy electronics of
the deformed graphene can be obtained from the modified Dirac
equation due to the modified hopping parameters from the
tight-binding model which are now a function of the atomic positions
$t(\textbf{r})$~\cite{neek2013}. The Dirac Hamiltonian in the
effective mass approximation in the presence of lattice deformation
(here out-of-plane deformation) introduces strain which induces an
effective gauge field $\vec{A}=\frac{2\beta_0\hbar}{3a_0
e}(\varepsilon_{xx}-\varepsilon_{yy},-2\varepsilon_{xy})$ where
$\beta_0$ ($\sim$2-3) is a constant~\cite{geim}. Using the strain
tensor components we found an analytical expression for $B$ as
function of $\theta$.
\begin{equation}
B=\frac{2\beta_0\hbar}{3a_0
e}(\varepsilon_{xx,x}-\varepsilon_{yy,x}+2\varepsilon_{xy,y}).\label{B}
\end{equation}
 We plot the height deformation ($h$) in
units of $h_0$ in Fig.~\ref{figB}(a) and the corresponding
pseudo-magnetic field in units of $h_0^2$ for $\theta=1.59^o$ in
Fig.~\ref{figB}(b). The pseudo-magnetic field has three-fold
symmetry and it is surprising that inside each MP unit cell the
field vanishes at the position of the extrema in the height
deformation (see the circles in  Fig.~\ref{figB}). It is also
worthwhile to mention that our study realizes in a natural way the
proposal for triaxial stress creation in graphene proposed by F.
Guinea \emph{et al.}~\cite{NatPhys} by using twisted graphene
sheets.
\section{summary}
In summary, we presented a theory for the out-of-plane deformation
of a twisted graphene sheet due to the vdW interaction with a
graphene substrate. By defining the MP structure, for the
out-of-plane deformation, as the minimum energy configuration we
derive an analytic solution without any fitting parameters. We found
excellent agreement for the height variation with our STM data for
different twist angles.

~~~~

 \emph{\textbf{Acknowledgment}}: This work was supported by the
Flemish Science Foundation (FWO-Vl) and the Methusalem Foundation of
the Flemish Government. M.N.-A was supported by the EU-Marie Curie
IIF postdoc Fellowship 299855. P.M.T. is thankful for the financial
support of the Office of Naval Research under Grant No.
N00014-10-1-0181 and the National Science Foundation under Grant No.
DMR-0855358. L.O.N. acknowledges the support of the American Society
for Engineering Education and Naval Research Laboratory Postdoctoral
Fellow Program. Work at the U.S. Naval Research Laboratory is
supported by the Office of Naval Research.

\section{appendix}
 The in-plane displacement vector for micron size graphene flake can be
written as
\begin{equation}
\vec{U}=\sum_{m,n}\vec{u}(\vec{r}+\vec{T}_{m,n}(\theta)),
\end{equation}
where the summation is taken over all MP unit cells and inside each
MP unit cell (see circles in Fig.~\ref{figB}) one can write
$\vec{u}(\vec{r},\theta)=C(\theta)(2xy,x^2-y^2)$ are the in-plane
components of the strain tensor, $\vec{T}_{m,n}(\theta)$ is the
translation vector of the MP lattice, and $C$ is a twist angle
dependent variable which determines the strength of the in-plane
strain. The corresponding in-plane strain elements written for each
MP unit cell are given by
\begin{eqnarray}
\varepsilon_{xx}=2Cy,\varepsilon_{yy}=-2Cy,\varepsilon_{xy}=2Cx\label{straininplane}
\end{eqnarray}
and the corresponding pseudo-magnetic field is a function of twist
angle but independent of position, i.e. $\frac{16C\beta_0\hbar}{3a_0
e}$. The corresponding principal axes are independent of twist
angle, i.e. the most tensile and compression directions in graphene
are $\frac{1}{2}tan^{-1}(\frac{-x}{y})$ and
$\frac{1}{2}tan^{-1}(\frac{y}{x})$, respectively, e.g. along   $y=x$
line two angles are $\pm\pi/8$.


\begin{thebibliography}{15}
\bibitem{Lopes} J. M. B. Lopes dos Santos, N. M. R. Peres, and A. H.
Castroneto, Phys. Rev. Lett \textbf{99}, 256802 (2007); Wen-Yu He,
Zhao-Dong Chu, and Lin He, Phys. Rev. Lett \textbf{111},  066803
(2013); E. Su\'{a}rez Morell, M. Pacheco, L. Chico, and L. Brey,
Phys. Rev. B \textbf{87}, 125414 (2013)
%
\bibitem{6}    K. Watanabe,    T. Taniguchi,   P. Jarillo-Herrero,     P. Jacquod, and B. J. LeRoy,
Nature Phys. \textbf{8}, 382 (2012).
%




\bibitem{prl}I. Brihuega, P. Mallet, H. González-Herrero, G. Trambly de Laissardière, M. M. Ugeda, L. Magaud, J. M. Gómez-Rodríguez, F. Ynduráin, and J.-Y. Veuillen, Phys. Rev. Lett. \textbf{109}, 196802 (2012).
\bibitem{eva} Eva Y Andrei, G. Li, and X. Du, Rep. Prog. Phys. \textbf{75}, 056501
(2012).

\bibitem{zhao} Zhao-Dong Chu, Wen-Yu He, and Lin He
Phys. Rev. B \textbf{87}, 155419 (2013); Wei Yan, Mengxi Liu,
Rui-Fen Dou, Lan Meng, Lei Feng, Zhao-Dong Chu, Yanfeng Zhang,
Zhongfan Liu, Jia-Cai Nie, and Lin He Phys. Rev. Lett. \textbf{109},
126801 (2012); Taisuke Ohta, Jeremy T. Robinson, Peter J. Feibelman,
Aaron Bostwick, Eli Rotenberg, and Thomas E. Beechem Phys. Rev.

\bibitem{PRL2008}J. Hass, F. Varchon, J. E. Millán-Otoya, M. Sprinkle, N. Sharma, W. A. de Heer, C. Berger, P. N. First, L. Magaud, and E. H. Conrad, Phys. Rev. Lett. \textbf{100}, 125504 (2008).

\bibitem{science2009} X. Li, W. Cai, Jinho An, S. Kim, J. Nah, D. Yang,
R. Piner, A. Velamakanni, I. Jung, E. Tutuc, S. K. Banerjee, L.
Colombo, R. S. Ruoff, Science \textbf{324}, 1312 (2009).

\bibitem{PHYSREVB2009} M. Ostler, I. Deretzis, S. Mammadov, F. Giannazzo, G. Nicotra, C. Spinella, Th. Seyller, and A. La Magna,
 Phys. Rev. B. \textbf{88}, 085408 (2013).


\bibitem{SACH} G. Giovannetti, P. A. Khomyakov, G. Brocks, P.
J. Kelly, and J. van den Brink, Phys. Rev. B \textbf{76}, 073103
(2007).

\bibitem{PRBref} S. Heinze, S. Blügel, R. Pascal, M. Bode, and R. Wiesendanger, Phys. Rev. B \textbf{58}, 16432
(1998).

\bibitem{spanu} L. Spanu, Sandro Sorella, and Giulia Galli, Phy. Rev. Lett \textbf{103}, 196401 (2009).

\bibitem{vdw} A. K. Geim and  I. V. Grigorieva,  Nature (London) \textbf{499},
419 (2013).


\bibitem{scientificreports} X. Chen, F. Tian, C. Persson, W Duan, and N-xian
Chen, Scientific Reports \textbf{3}, 3406 (2013).

\bibitem{PCCP} Irina V. Lebedeva, Andrey A. Knizhnik, Andrey M. Popov,
Yurii E. Lozovikda and Boris V. Potapkin, Phys. Chem. Chem. Phys.
\textbf{13}, 5687, (2011).
\bibitem{more} P. Xu, M. L. Ackerman, S. D. Barber, J. K. Schoelz, P. M. Thibado, V. D. Wheeler, L. O. Nyakiti, R. L. Myers-Ward, C. R. Eddy, Jr., and D. K. Gaskill
, Surface Science \textbf{617}, 113 (2013).


\bibitem{prbneek2012}M. Neek-Amal and F. M. Peeters, Phys. Rev. B \textbf{85}, 195445
(2012).
\bibitem{4} S. Tang, H. Wang, Y. Zhang,  A. Li, H. Xie, X.
Liu, L. Liu,  T. Li, F. Huang, X. Xie, and M. Jiang, Scientific
Reports \textbf{3}, 2666 (2013).
\bibitem{APL2014} M. Neek-Amal and F. M. Peeters, Appl. Phys. Lett. {\bf 104}, 041909 (2014).

\bibitem{Wallbank} J. R. Wallbank, A. A. Patel, M. Mucha-Kruczynski, A. K. Geim, and V. I. Fal`ko, Phy. Rev. B {\bf87}, 245 (2013).



%

\bibitem{neek2013}M. Neek-Amal, L. Covaci, Kh. Shakouri, and F. M.
Peeters, Phys. Rev. B \textbf{88}, 115428 (2013); M. Neek-Amal, L.
Covaci, and F. M. Peeters, Phys. Rev. B \textbf{86}, 041405 (2012);
M. Neek-Amal and F. M. Peeters Phys. Rev. B \textbf{85}, 195446
(2012).

%
%



%
%

%
%
%
%
%
%

\bibitem{geim}A. H. Castro Neto, F. Guinea, N. M. R. Peres,
 K. S. Novoselov, and A. K. Geim, Rev. Mod. Phys.  \textbf{81}, 109–162 (2009).


\bibitem{NatPhys} F. Guinea, M. I. Katsnelson, and A. K. Geim, Nat.
Phys. \textbf{2}, 31 (2010).






%




%
%
%
%
%


\end{thebibliography}
\end{document}